\documentclass[aps,prl,twocolumn,superscriptaddress,showpacs]{revtex4}

\include{epsf}
\usepackage{graphicx}
 
\newcommand{\cvo}{CaVO$_3$}
\newcommand{\B}{$\vec{B}$}

\newcommand{\kvec}{$\vec{k}$}
\newcommand{\ie}{{\it i.e.}}
\newcommand{\eg}{{\it e.g.}}
\newcommand{\degc}{$^\circ$C}

\begin{document}
 
\title{Fermi Surface of 3d$^1$ Perovskite CaVO$_3$ Near the Mott 
   Transition}
 
\author{I. H. Inoue}

\affiliation{Correlated Electron Research Center (CERC), 
  AIST Tsukuba Central 4, Tsukuba 305-8562, Japan}

\affiliation{IRC in Superconductivity,
  University of Cambridge, Madingley Road, Cambridge CB3 0HE, UK}

\author{C. Bergemann}

\affiliation{Cavendish Laboratory, University of Cambridge, Madingley
  Road, Cambridge CB3 0HE, UK}

\author{I. Hase}

\affiliation{Nanoelectronics Research Institute, AIST Tsukuba Central
  2, Tsukuba 305-8568, Japan}

\author{S. R. Julian}

\affiliation{Cavendish Laboratory, University of Cambridge, Madingley
  Road, Cambridge CB3 0HE, UK}

\begin{abstract}
 We present a detailed de Haas van Alphen effect study of the perovskite 
\cvo, offering an unprecedented test of electronic structure 
calculations in a 3d transition metal oxide.  Our experimental and 
calculated Fermi surfaces are in good agreement -- but only if we ignore 
large orthorhombic distortions of the cubic perovskite structure. 
Subtle discrepancies may shed light on an apparent conflict between the low 
energy properties of \cvo, which are those of a simple metal, and 
high energy probes which reveal strong correlations that place \cvo\ on 
the verge of a metal-insulator transition.
\end{abstract}

\pacs{71.18.+y, 71.20.-b, 71.30.+h}

\maketitle

Transition metal oxides (TMOs) have for many decades
been a rich source of novel and intriguing electronic behaviour.
The narrow bands in the 3$d$ TMOs are particularly susceptible to
correlation effects, leading to phenomena as unexpected and diverse as
high temperature superconductivity \cite{bednorz,orenstein}, colossal
magnetoresistance \cite{edtokura}, and the recently discovered
``orbitons'' in LaMnO$_{3}$ \cite{saitoh,tokura}.
The Mott transition \cite{mott,imada} -- a metal-insulator transition (MIT) 
driven by electron-electron interactions -- lies at the heart of TMO
physics, and the 3$d^1$ system CaVO$_{3}$ is emerging as an important
model compound that lies just on the metallic side of this
transition \cite{inoue94,fukushima,shirakawa,inoue98a}.

While the general principles involved in the MIT are probably 
understood, the predictive power of existing theories is poor 
when applied to real materials.  
On the one hand, models of the MIT that explicitly tackle dynamical 
many-body interactions use generic Hamiltonians that ignore 
the effects of real crystal structures; on the other, single-particle 
self-consistent band structure calculations based on the Local Density 
Approximation (LDA), which have had the most success at predicting the 
electronic structure of real metals,  
fail spectacularly in insulating TMOs such as NiO which is predicted by the 
LDA to be metallic.
This problem is compounded by a lack of high quality data on the 
electronic structure of TMOs: for example until now the exclusive source 
of information on the Fermi surface geometry in TMOs has been angle 
resolved photoemission spectroscopy (ARPES), but this technique detects 
a much higher energy response than is observed in transport and 
thermodynamic measurements, and the results are frequently distorted by 
surface effects. 

CaVO$_3$ is one of a series of 3d$^1$ perovskites that straddle the 
MIT.  It is a metal, but large orthorhombic distortions of the cubic 
perovskite crystal structure and the spectroscopic 
properties of \cvo\ suggest that it lies very close to the MIT 
\cite{inoue94,fukushima,shirakawa,inoue98a}. 
Photoemission spectroscopy in CaVO$_{3}$ appears to show: (a) the
formation of Hubbard bands \cite{inoue95,morikawa,marcelo,makino} ---a
precursor of the insulating state arising from strong electron
correlation; and (b) a weaker spectral intensity at the
Fermi energy $E_{\rm F}$ than is predicted by band structure
calculations \cite{inoue95,maiti}.
From these and other measurements a broad picture is emerging of a 
striking conflict between low and high energy scales in \cvo, the 
former being revealed for example in specific heat measurements and 
suggesting that \cvo\ is a simple metal with small mass enhancements 
(normally a signature of weak electron-electron interactions), 
while the latter are probed by spectroscopic methods and show 
{\em strong} electron-electron interactions. 
The origin of this conflict is unclear.  It may be an experimental 
artifact arising from enhanced sensitivity of spectroscopy to the 
surface as opposed to the bulk, but extra care has been taken to 
minimise this \cite{maiti}.
An intriguing suggestion is that the conflict, and the breakdown of 
the LDA that it implies, can be explained 
by a strong interaction that is non-local in space but local in time, 
such as for example the unscreened Coulomb interaction.  
But for this to occur, we require either a breakdown of screening 
in a material that at dc-frequencies at least has a normal metallic 
conductivity, or else 
a new non-local interaction such as might be mediated by orbital 
fluctuations. 

Further experimental data are clearly needed, and this was the 
motivation for the present de Haas van Alphen (dHvA) effect study, which 
allows an unprecedented and detailed comparison between 
modern electronic structure calculations and an experimentally  
determined Fermi surface in a 3d TMO. 
It should be noted that LDA calculations have an excellent record 
predicting the 
Fermi surface shapes of even strongly correlated electron systems 
such as UPt$_3$ \cite{taillefer}, in
which electron-electron interactions are enormously strong.
This is probably because the Fermi volume is invariant 
under interactions \cite{luttinger}: 
volume conserving distortions of the 
Fermi surface require not only a $k$-dependent (i.e.\ spatially non-local) 
interaction \cite{footnote1} such as was invoked to explain the 
photoemission results, but also an anisotropic Fermi surface, so that 
states in different regions of the Fermi surface undergo different shifts 
in energy. 
 
\begin{figure}
  \centerline{\epsfxsize=7.5cm \epsfbox{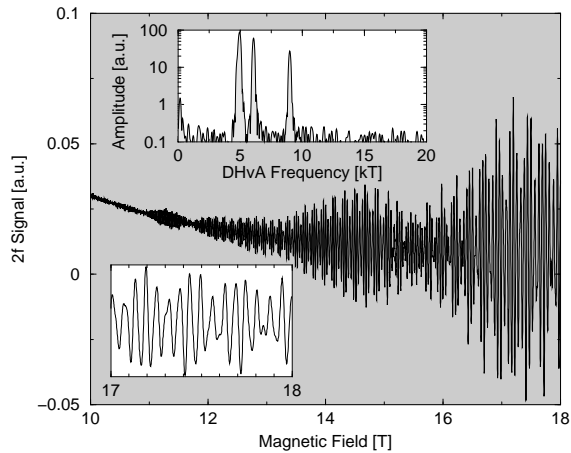}}
  \caption{%
    Example of a dHvA field sweep on \cvo\ with the applied field
    parallel to the $b$-axis. The lower inset is a blow-up of the
    high-field region.  The upper inset shows a logarithmic plot of
    the dHvA frequency spectrum, taken over the 18\,T to 15\,T field
    range.  }
  \label{dHvA}%
\end{figure}

Quantum oscillations, such as the
magnetisation oscillations in the de Haas-van Alphen (dHvA) effect, 
arise from the quantisation of cyclotron 
motion of charge carriers in a magnetic field.
This provides the most detailed and the most reliable information about
Fermi surface properties \cite{shoenberg}.
A prerequisite for the observation of the quantum oscillation effects
is that quasiparticles must
be able to complete a cyclotron orbit without scattering---this
typically requires mean-free paths of $\sim$\,1000\,\AA\ or more.
It is this requirement of very high purity single crystals that has
hitherto prevented quantum oscillations from being observed in $3d$
TMOs.

We used high-quality single crystals of \cvo, grown by the floating
zone method yielding single-crystalline grains of typical size
$\sim$\,$4 \times 4 \times 6$\,mm$^3$\@.
We kept the oxygen off-stoichiometry as small as possible by annealing
the sample in air at $\sim\!150$\degc\ for four days \cite{inoue94}.
The off-stoichiometry of Ca and V was below the detectability limit
(less than 1\,\%) of an inductively coupled plasma atomic emission
spectrometer.
The \cvo\ crystals are orthorhombic (P$_{\rm nma}$) containing four \cvo\
formulae in the unit cell with lattice parameters
$a = 5.314$\,\AA, $b=7.521$\,\AA, $c=5.339$\,\AA\ (Nakotte, H. \&
Jung, M.-H., private communication); the samples were oriented by Laue
x-ray diffraction.
%
\begin{figure*}
  \centerline{\includegraphics[scale=0.15]{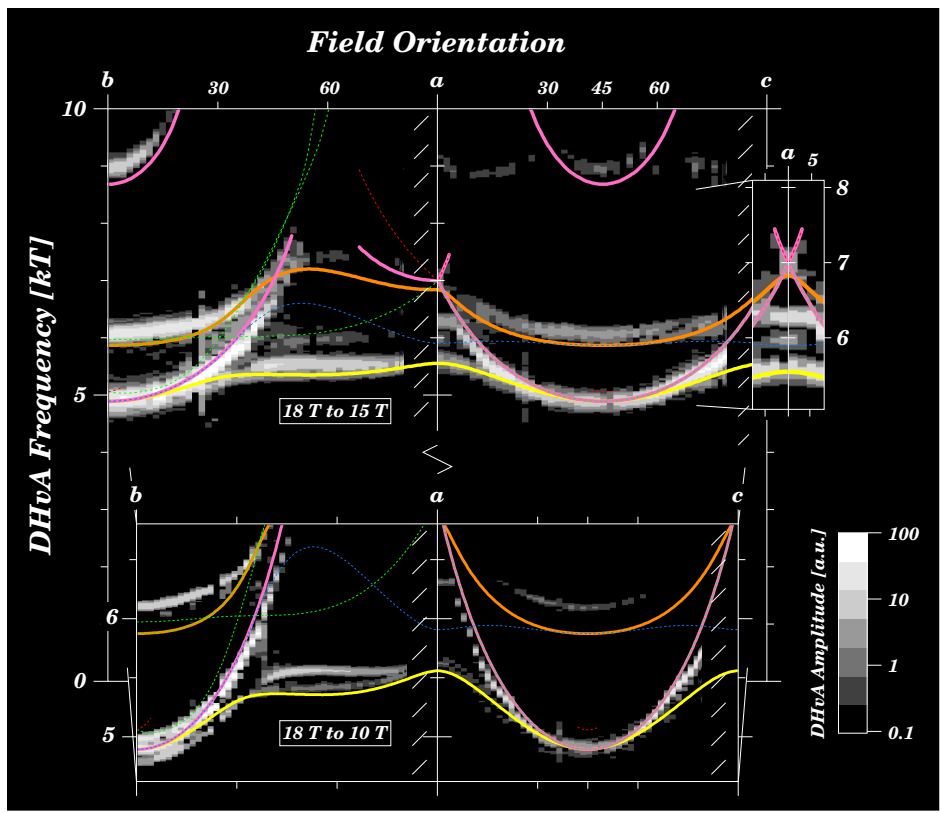}}
  \caption{%
    Density plot of the dHvA spectrum of \cvo\ as a function of the
    direction of \B\ (greyshaded)\@.  The main panel uses the
    high-field (18\,T $\to$ 15\,T) sections of each sweep, while the
    lower inset uses all available data (18\,T $\to$ 10\,T), with
    higher frequency resolution at the expense of signal-to-noise.
    The right inset shows a blow-up around the $a$-axis for fields in
    the $ac$-plane.  Superposed are the frequency predictions from LDA
    band structure calculations, for both the orthorhombic (thin
    dashed lines) and the hypothetical cubic (thick solid lines) unit
    cells---the colour code corresponds to
    Fig.~\protect\ref{surfaces}\@.  }
  \label{dplot}%
\end{figure*}
%

We have performed a thorough dHvA rotation study on two high-quality
single crystals of \cvo\ ($1 \times 1 \times 2$\,mm$^3$, residual
resistivity $\rho_0 \simeq 1.5\,\mu\Omega\,$cm) with the direction of
the applied magnetic field \B\ rotating in the ($a$-$b$) plane from 
$b$ to $a$, and in the ($a$-$c$) plane from $a$ to $c$. 
The experiments were carried out in a low-noise superconducting magnet
system in field sweeps from 18\,T to 10\,T, at temperatures down to  
50\,mK\@.
A modulation field of 14.3\,mT amplitude was applied to the sample,
and the second harmonic of the voltage induced at the pick-up coil
around the sample was recorded, essentially measuring $\partial^2
M/\partial B^2$.

A typical signal trace, demonstrating the high quality of our data,
can be viewed in Fig.~\ref{dHvA}. 
The Fourier transform as a function of $1/B$ gives the dHvA frequency
spectrum, with peaks at a frequency $F$ corresponding to extremal
(\ie, maximum or minimum) FS cross-sectional areas $S$ measured in
planes perpendicular to \B, via $S$\,=\,$2 \pi e F / \hbar$.

\begin{figure}
  \centerline{\epsfxsize=7.5cm \epsfbox{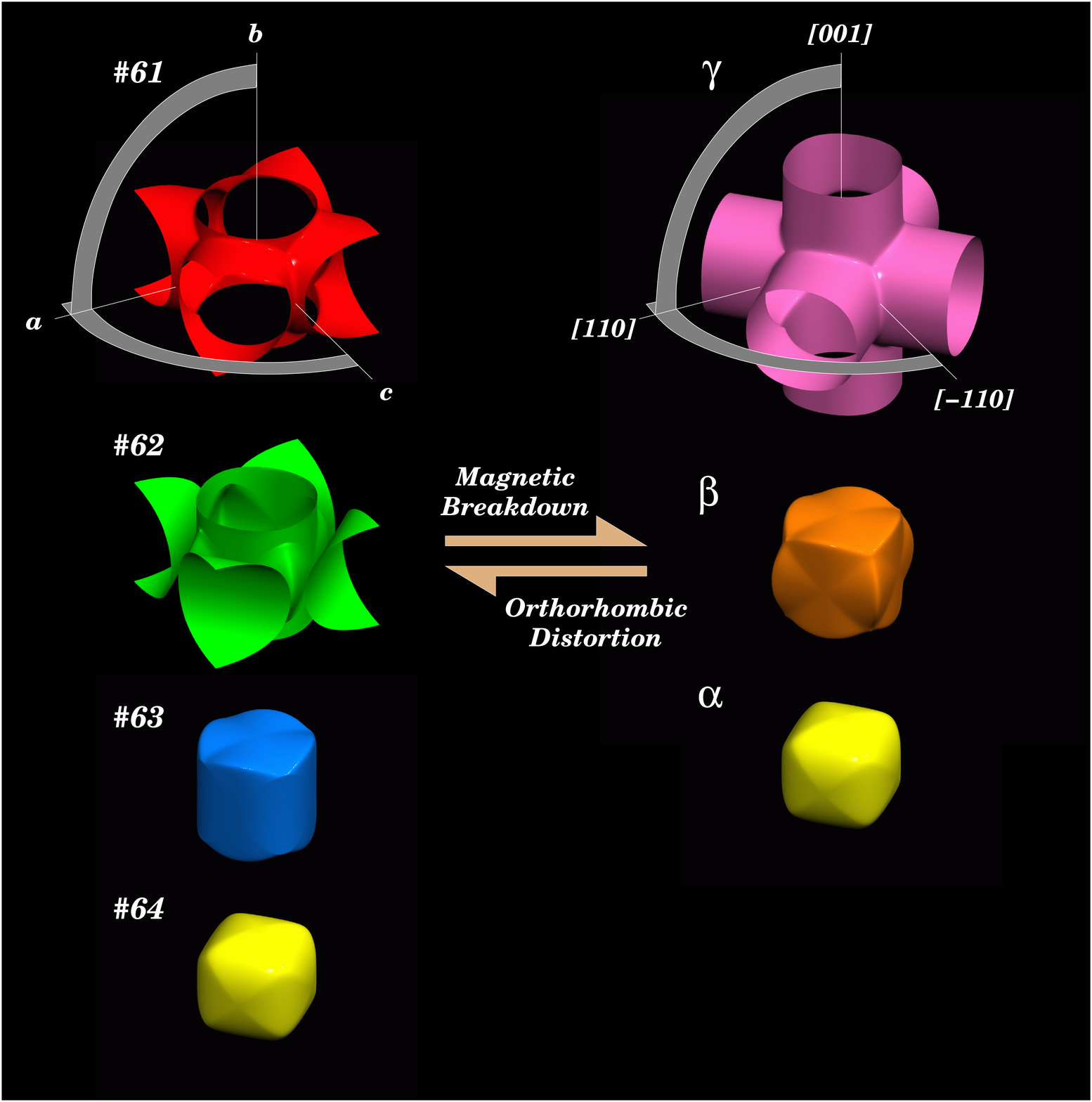}}
  \caption{%
       FS sheets predicted by LDA. The right panels show the FS sheets
       in a {\em cubic\/} TB fit to the LDA data, while on the left
       side these sheets are folded over into the orthorhombic BZ. The
       field directions in the rotation study are indicated around the
       surfaces in the top row.
     }
  \label{surfaces}%
\end{figure}

In Fig.~\ref{dplot}, we present a density plot of the dHvA spectrum as
a function of the direction of \B, representing the information of
about 100 field sweeps at angular intervals of 2$^\circ$\@.
In comparison, we have calculated the dHvA frequencies $F_{\rm calc}$
from four FS sheets (see Fig.~\ref{surfaces}) obtained by a band
calculation using the computer code KANSAI-94 with the LDA
scheme \cite{gunnarson} for the one-electron exchange-correlation
potential.
In the calculation, we used basis functions in
 $|\vec{k}+\vec{G}|$\,$<$\,$K_{\rm max}$\,=\,$6.20$\,(2$\pi$/\AA),
where \kvec\ is the wavevector and $\vec{G}$
is a reciprocal lattice vector, to obtain about 1440 basis
linear augmented plane waves; the core and valence states were
calculated self-consistently by the scalar-relativistic
scheme \cite{koelling77} with muffin-tin radii of $0.25\,a$ for Ca,
$0.20\,a$ for V, and $0.14\,a$ for O\@.

As seen in Fig.~\ref{dplot}, there are very significant
discrepancies between $F_{\rm calc}$ (thin dashed lines) and the
experimental dHvA frequencies $F_{\rm exp}$ (grey-shaded). To list a
few, (a) the $\sim 9$\,kT branches in $F_{\rm exp}$ have no
correspondence in $F_{\rm calc}$, and (b) the $^{\#}63$ sheet is
not observed in $F_{\rm exp}$ unless we accept a deviation of about
20\,\%. Only $^{\#}64$ and some of the orbits on $^{\#}62$ show
satisfactory coincidence.

Although we observe this large discrepancy between experiment and LDA,
we do not believe that this is the end of the story, because we have
found that excellent agreement between theory and experiment can be
regained {\em if we ignore the orthorhombic distortion in the crystal
  structure}, and recalculate the Fermi surface as though \cvo\ were
cubic.
We attempted to fit the calculated LDA bands of the
orthorhombic lattice structure to a hypothetical ``cubic''
tight-binding (TB) Hamiltonian consisting of V $3d_{xy}$, $3d_{yz}$
and $3d_{zx}$ orbitals, with nearest neighbour, next-nearest neighbour
and hybridization matrix elements.
Then, the results (thick solid lines in Fig.~\ref{dplot}) show
extremely good agreement with experiment: the differences in $F$ are only
4\,\% or less.
As depicted in Fig.~\ref{surfaces}, in this assumption, the orthorhombic
FS sheets $^{\#}61$, $^{\#}62$, and $^{\#}63$ will coalesce into the
hypothetical ``cubic'' FS sheets $\beta$ and $\gamma$, while the
$^{\#}61$ sheet remains unaffected.
The $\gamma$ sheet connects up with itself in the
neighbouring BZ, forming a ``jungle gym''-type structure.
The large hole orbit in this ``jungle gym'' explains the
high-frequency features in $F_{\rm exp}$, \eg\ along the $b$-axis.

\begin{figure}
  \centerline{\includegraphics[scale=0.5]{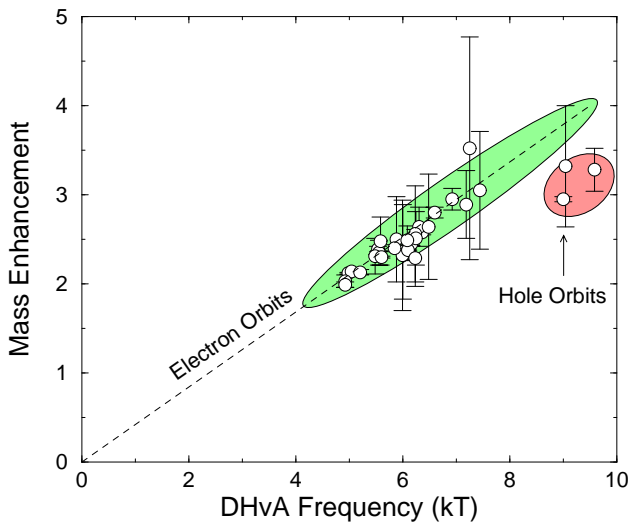}}
  \caption{
       The ratio of the effective masses to the free electron mass
       for all observed $F_{\rm exp}$ in \cvo\ at various field orientations,
       spaced by about 15$^\circ$ along $b \to a \to c$.  The
       masses scale with the dHvA frequency, except for the 9\,kT hole orbit.}
  \label{mass}
\end{figure}

We also deduced the effective quasiparticle masses $m^\ast$ from the
thermal broadening factor $R_T = X/\sinh X$, $X = 2\pi^2k_BT m^\ast/e\hbar
B$, by comparing dHvA amplitudes at
different temperatures in runs from 50\,mK up to 1.1\,K. As seen in
Fig.~\ref{mass}, we have $m^\ast \propto F$ for all orbits and for all
field directions assessed, save for the hole orbit of the ``jungle
gym''.
Such homogeneous scaling of $m^\ast$ with $F$ is seen in the TB model in
two-dimensional metals with a low-filling and with primarily
nearest-neighbour interactions. This results also supports our 
assumption of the effective cubic symmetry, because the 
FS sheets of a cubic $d^{1}$ system are essentially
formed from three intersecting cylinders containing the $d_{xy}$,
$d_{xz}$, and $d_{yz}$ states respectively, and each has a 
two-dimensional character.
The specific heat deduced from dHvA is 8.6\,mJ/mole\,K, which is in
good agreement with the bulk experimental value \cite{inouethesis} of
7.3\,mJ/mole\,K and is enhanced over the unrenormalized LDA density of
states by about a factor of two.
%

Although some details in the experimental dHvA spectrum of \cvo\ 
remain yet to be identified and are possibly associated with remnant
features of the intrinsic orthorhombic structure, the overall angle
dependence of the dHvA frequencies is altogether well captured by the
hypothetical ``cubic'' TB model, rather than the orthorhombic LDA 
band calculation.
  
Our apparently {\em ad hoc\/} assumption that the effective crystal {\em
symmetry\/} sensed by the $3d$ electrons is cubic, may be justified if the gaps
in the energy bands that are produced by the orthorhombic distortion
are small enough --- an effect known as magnetic breakdown
(MB) \cite{shoenberg}.
However, its appearance here seems to be surprising, since the
orthorhombic distortion in \cvo\ is not small (the buckling
of the V-O-V bond angle is 160$^\circ$). 

We hope that our results will inspire further, more detailed electronic 
structure calculations.  
It will be interesting to see if a high-resolution LDA calculation will 
show the small gaps due to the orthorhombic distortion that 
our results imply, and thereby reproduce the predictions of our 
tight binding model. 
If not, some more exotic explanation of the non-appearance of 
orthorhombic gaps may be required. 
Also, while the differences between our tight-binding Fermi surface 
and the measurements are small, they are systematic. In particular, it 
can be seen in the lower insert of Fig.~\ref{dplot} that the 
fit to the $\beta$ surface is clearly worse than the others.  
The $\beta$ sheet is composed of states that are all near an intersection 
of the three cylinders of the cubic Fermi surface model, so perhaps this 
is indeed a signature of a non-local interactions, but detailed 
computational work is clearly required to clarify this, starting with a 
high-resolution LDA calculation of the electronic structure.

In summary, this first comprehensive investigation of the dHvA effect
in a $3d$ TMO has unraveled the shape of the FS sheets and thus the
electronic structures near the
Mott transition in \cvo. 
The experimental FS cannot be explained by the LDA band calculation
for the intrinsic orthorhombic lattice, but the results are well 
captured by a hypothetical ``cubic'' electronic structure.
At low energy scales, \cvo\ is seen to be a standard metal with
moderate mass enhancement, in marked contrast to the strong
correlations visible at higher energies.
While further efforts are required to comprehend this
discrepancy and to gain a quantitative understanding of the origin of
the MB effects, this work provides at least a first glimpse of the
completely unexplored fermiology near the Mott transition.

\acknowledgements
We thank D. M. Broun, H. Bando, and W. Y. Liang for their support.
We also thank H. Nakotte and M.-H. Jung for sharing their data prior to
publication. This work was supported in part by the JST Overseas Research
Fellowship and the UK EPSRC.


\begin{thebibliography}{99}
\bibitem{footnote1} 
More technically, the self-energy $\Sigma(k,\omega)$ has to be 
$k$-dependent.  This is usually ignored in theories of the 
MIT, but it seems plausible that as screening breaks down on the 
verge of the MIT such a $k$-dependence could arise.
\end{thebibliography}
\end{document}